# Barometric Effect of High-Energy Muons ( > 230 GeV)

M.Berkova[1], V.Yanke[1], L.Dorman[1], V.Petkov[2], M.Kostyuk[2], R.Novoseltseva[2], Yu.Novoseltsev[2], P.Striganov[2], M.Boliev[2], V.Volchenko[2], G.Volchenko[2], A.Yanin[2], I.Dzaparova[2], M.Kochkarov[2]

[1]Institute of Terrestrial Magnetism, Ionosphere and Radiowave Propagation RAS of N.V. Pushkov (IZMIRAN), Moscow, Troitsk, RU-142190, Russia
[2]Instutute for Nuclear Research RAS, 60th Oct. Anniversary prospekt 7A, Moscow, RU-117 312, Russia

According to the Baksan underground scintillation telescope's data the barometric coefficients for six intervals of zenith angles of particle registration were obtained for ten years (2003-2012). The obtained barometric coefficients exceed an order of magnitude the theoretical values − the obtained value for the vertical $\beta$ = (-0.051 ± 0.010) %/mb. Nevertheless, these results are in good agreement with the data obtained for the barometric effect of high-energy muons in a number papers of other authors. And there is a noticeable increase of the absolute values of the measured barometric coefficients with increase of the threshold energy of muons.

E-mail: qvark8@yandex.ru, vpetkov@inr.ru, yanke@izmiran.ru

**The Baksan underground scintillation telescope**

The Baksan Underground Scintillation Telescope (BUST) of the Institute for Nuclear Research RAS is located in the Caucasus mountains (43.28° N and 42.69° E) inside the mine (dimensions 24×24×16 m$^3$ ) under the Andyrcha mountain. There is 350 m of rock vertical depth (850 mwe) over the telescope. The BUST is parallelepiped (16.7×16.7×11.1 m$^3$ ) consisting of 8 planes (4 inner and 4 outer) of detectors (each of 70×70×30 cm$^3$) with liquid scintillater (Figure 1). The observatory level is 1700 m above sea level, average pressure is 820 mb. The muon threshold energy is 220 GeV. But because of the complex topography over the installation (according to angles and directions) one can obtain results to a depth > 6000 g/cm$^2$ [Alekseev et al., 1980]. The BUST's experimental data cover a wide range of threshold energies (0.2 - 10 TeV) and zenith angles (0° - 90°). Data for angles θ>70° are characterized by low intensity and therefore - the small statistics.

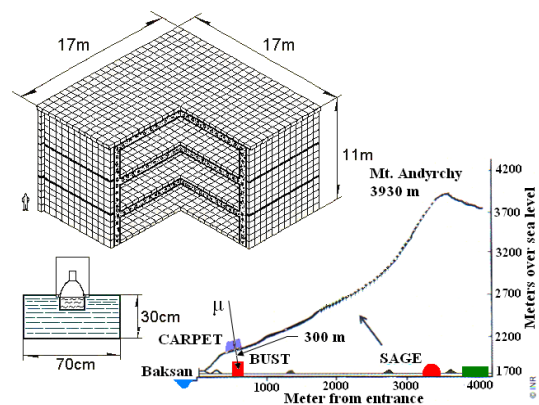

**Fig. 1**. The Baksan Underground Scintillation Telescope

The BUST's data are integrated over the azimuthal angles and presented by six intervals of zenith angles (vertical v0 (0°<θ <34°); interval 01 (34°<θ<48°); interval 02 (48°<θ<60°); interval 03 (60°<θ<71°); interval 04 (71°<θ<80°); interval 05 (80°<θ<90°)). The average hourly count rates for each interval of zenith angles are presented in Table 1,

**Table 1.** The average hourly BUST's count rates for six intervals of zenith angles.

| total | v0<br>0°<θ<34° | 01<br>34°<θ<48° | 02<br>48°<θ<60° | 03<br>60°<θ<71° | 04<br>71°<θ<80° | 05<br>80°<θ<90° |
|---|---|---|---|---|---|---|
| 44267 | 13521 | 11524 | 9765 | 7006 | 2203 | 251 |

Because of the complex topography over the telescope each interval has a wide range of threshold energies. The main contribution to the count rate (~ 80%) comes from zenith angles $0^o$ - $60^o$ that approximately corresponds to the range of the effective threshold energies from 0,2 to 0,3 TeV. And for zenith angles $50^o$ - $70^o$ the threshold energies lies in a very large range - from 0,5 to 10 TeV [Voevodsky et al., 1993].

A calculation of the BUST's muon energy spectrum was made in a preprint P-0379 [Gurentsov, 1984]. According to this work a dependence of the threshold energies of the registered particles on the azimuthal angles for the full range of zenith angles (from $10^o$ to $85^o$) are built In Figure 2. The curves (in steps of $5^o$) are grouped by the presented intervals of zenith angles: vertical v0 ($0^o<\theta<34^o$) corresponds to the red colour; the interval 01 ($34^o<\theta<48^o$) – green; the interval 02 ($48^o<\theta<60^o$) –orange; the interval 03 ($60^o<\theta<71^o$) – blue; the interval 04 ($71^o<\theta<80^o$) – grey; the interval 05 ($80^o<\theta<90^o$) – brown. As in the preprint P-0379 data are given on the zenith angles in steps of $5^o$ then each of the six intervals (except the interval 05) corresponds to two or more curves. It is clearly seen in Figure 2 that threshold energies vary by orders of magnitude within one interval for angles $\theta > 45º$ (depending on the amount of the rock). This fact complicates determination of the effective threshold energy for each of the intervals considered.

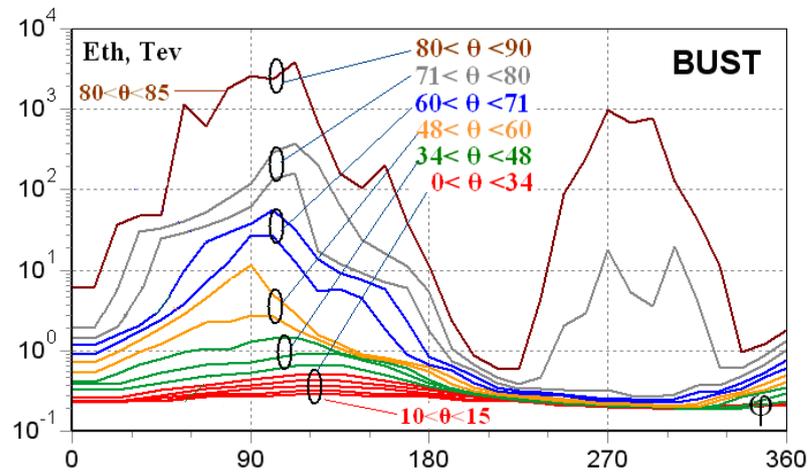

**Fig. 2.** Dependence of the threshold energies on azimuthal angles for the full range of zenith angles in steps of $5^o$ for the BUST: vertical v0 ($0^o<\theta<34^o$) - red curves; the interval 01 ($34^o<\theta<48^o$) – green curves; the interval 02 ($48^o<\theta<60^o$) – orange curves; the interval 03 ($60^o<\theta<71^o$) – blue curves; the interval 04 ($71^o<\theta<80^o$) – grey curves; the interval 05 ($80^o<\theta<90^o$) – a brown curve

Nevertheless in this paper the following values of the effective threshold energies were adopted for each of the zenith angle intervals: a) for the vertical v0 ($0^o<\theta<34^o$) – $E_{th}$ ~ 230 GeV; b) for the interval 01 ($34^o<\theta<48^o$) – $E_{th}$ ~ 280 GeV; c) for the interval 02 ($48^o<\theta<60^o$) – $E_{th}$ ~ 380 GeV; d) for the interval 03 ($60^o<\theta<71^o$) – $E_{th}$ ~ 850 GeV; e) for the interval 04 ($71^o<\theta<80^o$) – $E_{th}$ ~ 1700 GeV; f) for the interval 05 ($80^o<\theta<90^o$) – $E_{th}$ ~ > 7000 GeV (see Table 2).

**Table 1**. Effective threshold energies for the intervals of zenith angles.

| Интервал | θ | $E_{th}$, GeV |
|---|---|---|
| **v0** | $0^o<\theta<34^o$ | 230 |
| **01** | $34^o<\theta<48^o$ | 280 |
| **02** | $48^o<\theta<60^o$ | 380 |
| **03** | $60^o<\theta<71^o$ | 850 |
| **04** | $71^o<\theta<80^o$ | 1700 |
| **05** | $80^o<\theta<90^o$ | > 7000 |

**Barometric effect as observed by underground detectors.**

The barometric effect is well studied within the framework of the meteorological effects of secondary cosmic ray component theory, and it is described in detail in [Dorman, 1972]. According to the theory, it is assumed that the barometric effect of high energy muons is negligible, as its absolute value decreases with increasing the threshold energy of the registered particles. In the book [Dorman, 1972] the barometric coefficients for high-energy muons obtained by different authors for counter telescopes at various depths are presented. Their values are in good agreement with the theory, i.e. with the depth (or energy) increasing the absolute value of the barometric coefficient decreases.

But J.E.Humble et al. in 1979 and P.R.A.Lyons et al. in 1981 in [Humble et al., 1979] and [P.R.A.Lyons et al., 1981] tried to explain an unusually large value of the barometric coefficient observed at Poatina, Tasmania (357 m.w.e. in depth, $E_{th} > 100$ Gev) which was about 7 times more than the theoretically predicted one. Thus according to the Poatina's data from 1972 to 1976, taking into account only the pressure change, the barometric coefficient β = (-0,047 ± 0,0016) %/mb was obtained by using linear regression, while a theoretical value was (-0,007) %/mb. In the paper [Humble et al., 1979] it was also reported about large seasonal variations of the barometric coefficient. To explain this result an assumption was made that the barometric coefficient is also affected by unaccounted parameters of the upper atmosphere (100 mb and above). To check this assumption, in the paper a multiple regression analysis of the data was carried out over the same period taking into account all three parameters – pressure at the observation level, altitude and temperature of the 100 mb layer (information about the higher layers were not available). As a result, the barometric coefficient was obtained β = (-0,0501 ± 0,0026) %/mb, that was not too different from the previous value. The authors concluded that the discrepancy of the observed and theoretical barometric coefficients may be caused by unaccounted muons from kaons, as well as the unaccounted impact of temperature of the upper atmosphere layers above 100 mb, where muons of such high energies are mostly produced.

In a paper [Lyons et al., 1981] a significant negative correlation between pressure and temperature of the upper atmosphere was defined. Considering this result and the possible temperature errors, the authors evaluated the barometric coefficient for Poatina and got β in the range from (-0,0214 ± 0,0020) to (-0,0257 ± 0,0019) %/mb. These values are two times less than those obtained earlier, but still several times higher than the theoretical ones. In a paper [Yasue, 1981] it was reported that for the underground telescope Matsushiro (250 m.w.e. in depth) a barometric coefficient was obtained as β = (-0,045 ± 0,005) that is five times more than the theoretically expected value. It was explained by the temperature effect too.

Later S.Sagisaka in a work [Sagisaka, 1986] presented barometric coefficients for a large number of underground detectors at different depths. All the data presented in this paper were in good agreement with theoretical calculations. But, for the deepest detectors Matushiro (250 mwe in depth) and Poatina (365 mwe in depth) the barometric coefficients were significantly more than theoretically calculated values: β = (-0,027 ± 0.004) %/mb for Matushiro [S. Sagisaka, 1983] and β = (-0.047±0.002) %/mb for Poatina [Fenton, 1975], [Humble, 1979], [Lyons, 1981]. S.Sagisaka explained this discrepancy as the possible influence of the temperature effect, because observed barometric coefficients were obtained by simple correlations between the variations of the measured muon intensity ΔI and the pressure ΔP, without taking into account temperature changes.

It was reported about a large value of the barometric coefficient for groups of muons (the registered events corresponded to the primary particles energies of ~ $10^{15}$ eV) determined at the DECOR installation (Moscow Engineering Physics Institute), unlike the barometric coefficient for single muons [Tolkacheva, 2011]. With the double linear regression (taking into account variations of pressure and temperature at the same time) the barometric coefficient for groups of muons was obtained as β=(-0.314±0.002) %/mb. The observed effect was explained by a change of the spatial distribution function for muons in extensive air showers.

Let us recall that for single muons (at the Earth's surface, $E_{th}$~0.4GeV) barometric coefficient is approximately equal to β≅-0.15 %/mb. And according to the theory of meteorological effects [Dorman, 1972], the absolute value of the barometric coefficient should be decreased with increasing energy of the registered particles.

**Barometric effect as observed by the BUST**

The gauge of the absolute atmospheric pressure at the Baksan Neutrino Observatory is built on a semiconductor detector (Motorola MPX4115A). It is not precise. Therefore, to make sure that the pressure sensor was sufficient for our task accuracy, we estimated the barometric coefficient of the neutron component by the Baksan neutron monitor data. This can be done via the Internet project [Kobelev, 2011; Paschalis, 2012]. There is the barometric coefficient calculation for the Baksan neutron monitor for 2010 in Fig. 3. According to the result, it can be concluded that the pressure sensor data can be used with confidence to assess the barometric coefficient of the muon component registered by the BUST.

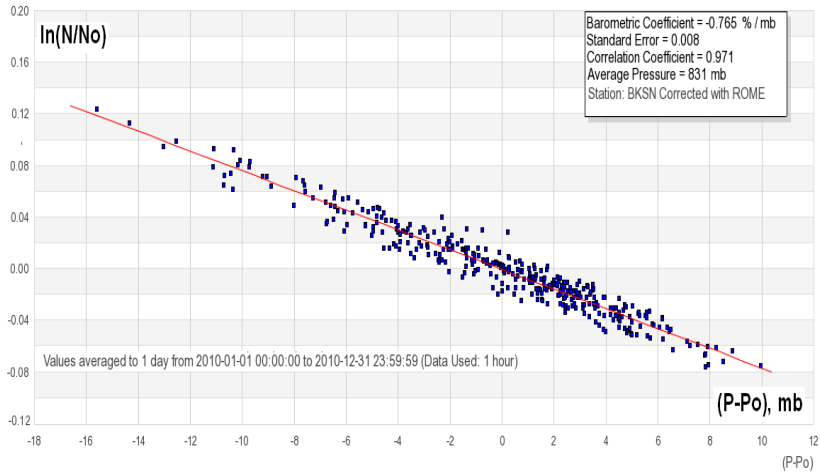

**Fig. 3.** Correlations between count rate of the Baksan neutron monitor and pressure for 2010. The data were corrected for the variations of the reference station Rome.

In a paper [Andreyev et al., 1987] the barometric effect for such depths as the BUST was estimated as too small value (β~-0,004 %/mb), therefore it could be neglected. For the same reason the barometric effect was not considered in a later paper [Kozyarivsky et al., 2004]. In the study of the temperature effect for the BUST [Berkova et al., 2011] according to the total count rate over all directions for 2009-2010, the authors also assumed that the barometric effect for high-energy muons (> 200 GeV) is negligible.

However, before starting the temperature effect analysis for a ten-year period (from 2003 to 2012) of observations (for six channels - vertical and five intervals of zenith angles), it was decided to release the monitoring data even from a small barometric effect.

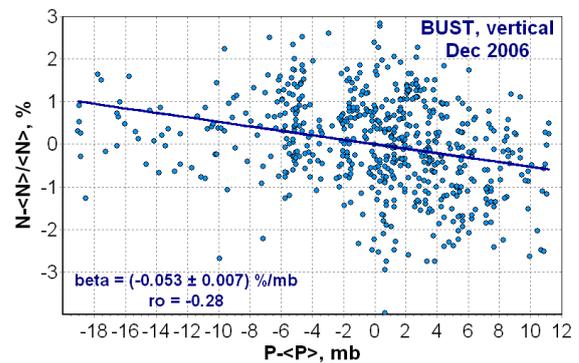

**Fig. 4.** Correlation between the hourly telescope's data (vertical) and pressure for December 2006.

To determine the barometric coefficient, several months with the highest amplitude of pressure changes (~ 20 mb) were selected. Using the hourly data, barometric coefficients for all the intervals of zenith angles were obtained for each selected month by a simple linear regression. The correlation between the hourly telescope's data (vertical) and pressure for December 2006 is presented in Fig. 4.

The average values of the barometric coefficients and correlation coefficients (for the vertical and for five intervals of zenith angles) obtained by using the hourly data for the selected months are presented in Table 3 (version I). The obtained experimental barometric coefficients exceed an order of magnitude or more the theoretical ones.

**Table 3.** The average barometric and correlation coefficients (for the vertical and for five intervals of zenith angles) obtained in three ways for 2003-2012.

| | channel | v0 $0°<\theta<34°$ | 01 $34°<\theta<48°$ | 02 $48°<\theta<60°$ | 03 $60°<\theta<71°$ | 04 $71°<\theta<80°$ | 05 $80°<\theta<90°$ |
|---|---|---|---|---|---|---|---|
| | $E_{th}$, GeV | 230 | 280 | 380 | 850 | 1300 | > 7000 |
| | d, m.w.e. | 940 | 1170 | 1400 | 1980 | 3260 | 7000 |
| | | For the selected months (by hourly data) | | | | | |
| I | β %/mb | -0,044 (±0,008) | -0,043 (±0,008) | -0,048 (±0,009) | -0,049 (±0,010) | -0,067 (±0,017) | -0,080 (±0,051) |
| | ρ | -0,23 | -0,22 | -0,22 | -0,19 | -0,16 | -0,06 |
| | | For average annual data | | | | | |
| II | β %/mb | -0.061 (±0.031) | -0.075 (±0.019) | -0.124 (±0.033) | -0.089 (±0.015) | -0.063 (±0.019) | 0.078 (±0.084) |
| | ρ | -0.63 | -0.85 | -0.84 | -0.93 | -0.80 | 0.35 |
| | | For all months (by hourly data) | | | | | |
| III | β %/mb | -0.051 (±0.010) | -0.044 (±0.010) | -0.048 (±0.010) | -0.049 (±0.015) | -0.069 (±0.020) | -0.157 (± 0.030) |

Analyzing the data obtained, it is clear that the more statistics the better correlation, and vice versa. It is difficult to say about any correlation for the largest angles, as their count rate is too small. In simple linear regression the model used to describe the relationship between two variables (in our case – telescope's count rate and pressure), but at the same time, one can not ignore the influence of unaccounted temperature effect. With that, for the hourly data on a monthly interval there are always diurnal and semi-diurnal temperature variations [Humble, 1979], [Lyons, 1981].

To avoid a possible influence of the temperature effect on the correlation between pressure and the telescope's data the linear regression was performed on the average annual data for 10 years. In this approach the influence of daily and annual temperature variations was excluded as they are averaged over such a long interval. Figure 5 shows the correlations between annual averages of pressure and average annual BUST's data for all six intervals of zenith angles for 2003-2012.

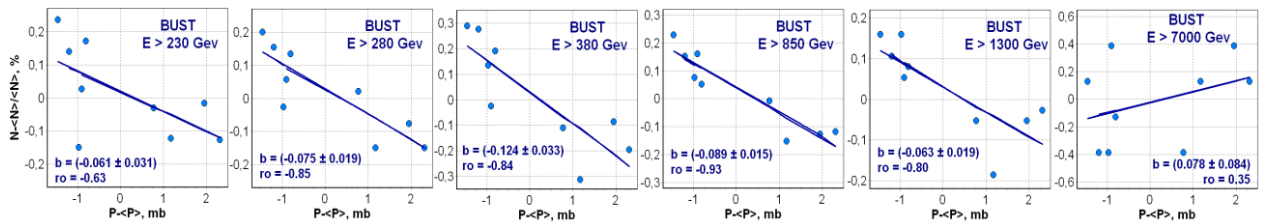

**Fig. 5.** Correlations between annual averages of pressure and average annual BUST's data for all six intervals of zenith angles for 2003-2012

The barometric coefficients obtained by the correlation analysis of average annual count rate for each direction and the corresponding correlation coefficients are presented in Table 3 (version II).

To minimize errors due to low statistics in the case of selecting several months, work to determine the barometric coefficients for hourly data for all 120 months under consideration (for 2003-2012) was carried out. Because of the data lack the barometric coefficient was not obtained only for January 2004. For the rest 119 months the barometric coefficients for each of the six intervals of zenith angles considered were obtained by the hourly data with varying statistical confidence (because of the problems with the data). The results with a positive correlation and with a correlation coefficient less than 0.1 were excluded from the values obtained. The remaining values were averaged with the weights as

$$\beta = \sum_{i=1}^{n} \frac{\beta_i}{\sigma_i^2} / \sum_{i=1}^{n} \frac{1}{\sigma_i^2} . \qquad (1)$$

where $\sigma_i$ – standard deviation of the barometric coefficient values $\beta_i$ obtained with the linear regression. Averaged values of the barometric coefficients (by hourly data) for all intervals of zenith angles for all epochs 2003-2012 are presented in Table. 4. And the average barometric coefficients from Table 4 are also presented in Table 3 (version III).

**Table 4.** Averaged values of the barometric coefficients (by hourly data) for all intervals of zenith angles for all epochs 2003-2012

| channel | v0 $0°<\theta<34°$ | 01 $34°<\theta<48°$ | 02 $48°<\theta<60°$ | 03 $60°<\theta<71°$ | 04 $71°<\theta<80°$ | 05 $80°<\theta<90°$ |
|---|---|---|---|---|---|---|
| $E_{th}$, GeV | 230 | 280 | 380 | 850 | 1300 | >7000 |
| d, m.w.e. | 940 | 1170 | 1400 | 1980 | 3260 | 7000 |
| Year | $\beta_{v0}$, %/mb | $\beta_{01}$, %/mb | $\beta_{02}$, %/mb | $\beta_{03}$, %/mb | $\beta_{04}$, %/mb | $\beta_{05}$, %/mb |
| 2003 | -0.052 | -0.041 | -0.045 | -0.047 | -0.087 | -0.166 |
| 2004 | -0.043 | -0.045 | -0.048 | -0.043 | -0.050 | -0.180 |
| 2005 | -0.045 | -0.046 | -0.053 | -0.059 | -0.069 | -0.116 |
| 2006 | -0.064 | -0.050 | -0.073 | -0.051 | -0.090 | -0.133 |
| 2007 | -0.057 | -0.032 | -0.043 | -0.055 | -0.054 | -0.194 |
| 2008 | -0.039 | -0.040 | -0.044 | -0.045 | -0.059 | -0.103 |
| 2009 | -0.054 | -0.042 | -0.052 | -0.043 | -0.071 | -0.165 |
| 2010 | -0.037 | -0.052 | -0.041 | -0.046 | -0.067 | -0.188 |
| 2011 | -0.071 | -0.050 | -0.045 | -0.054 | -0.096 | -0.160 |
| 2012 | -0.048 | -0.044 | -0.032 | -0.050 | -0.045 | -0.169 |
| average 2003-2012 | -0.051 (± 0.010) | -0.044 (± 0.010) | -0.048 (± 0.010) | -0.049 (± 0.015) | -0.069 (± 0.020) | -0.157 (± 0.030) |

All the results of the Table 4 are shown in Figure 6 as a dependence of the average values of the barometric coefficients on the energy corresponding to the intervals of zenith angles under consideration. The average values are depicted in Fig. 6 by thin lines, and a thick line (with markers showing the values of the muon energy) corresponds to the average values for the entire period of 10 years.

It is interesting that up to TeV energies the barometric coefficient within the errors does not change. However, after $10^{12}$ eV, there is a strong increase of the absolute value of the barometric coefficient, which may indicate of a qualitative change in the processes accompanying the muons origination.

**Results and discussion**

It is seen from Table 3 (version I and II) that the correlation coefficients ρ for all directions for the annual data are much higher than for the hourly ones. This may confirm the assumption that the temperature affects the barometric coefficients obtained by linear regression for hourly data. At the same time the barometric coefficients turned out to be much larger.

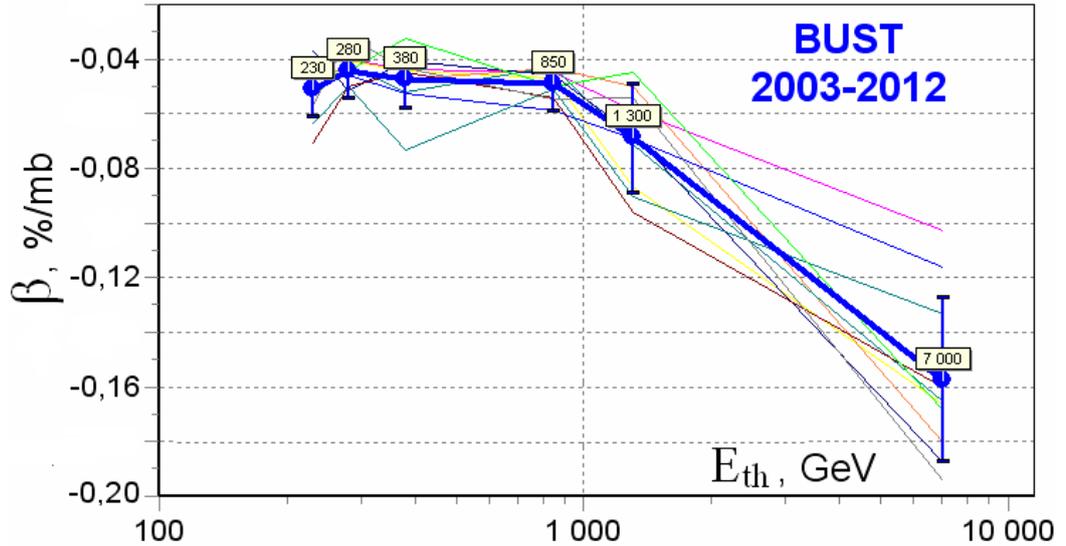

**Fig. 6.** Dependence of the barometric coefficients on the energy, obtained by the hourly data for 2003-2012 (thin lines). The markers show the values of the muon energy for all six intervals of zenith angles, and correspond to the values averaged for 10 years.

Significantly smaller errors for the hourly data can be noted, as well as the absolute value of the barometric coefficient increases with the energy increasing. For the barometric coefficients obtained by the annual data, the errors are much larger, and there is a break of the curve for the energy > 850 GeV. The difference in the magnitude of errors can be explained by the fact that the correlation analysis for the hourly data was carried out for many more points than for the annual data. But for all that, if we ignore the last point ($E_{th}$ > 7000 GeV) as the most unreliable due to the small statistics, it may be noted that the barometric coefficients obtained by the hourly and the annual data are of the same order, and more than an order of magnitude higher than the theoretically predicted values of the barometric coefficients for these energies, as it follows from the theory [Dorman, 1972] and [Sagisaka, 1986].

To combine all results obtained and referred to, a figure from [Sagisaka, 1986] was extended (Fig. 7). The barometric coefficients for hourly data for all the intervals of zenith angles (red points 15-20) for the threshold energies (from 230 to 7000 GeV) obtained in this work have been added in Fig. 7. The points from 0 to 14 and the calculated curves are from [Sagisaka, 1986]. The energy scale in Fig. 7 has been extended right up to the energies of the primary particles of EAS (up to $10^7$ GeV) that allowed to plot barometric coefficients for EAS (black triangles 21-27) [Dorman, 1972]. A red point 28 corresponds to the barometric coefficient obtained in [Tolkacheva, 2011] for groups of muons. For the Andyrchi air shower array (energies from 2 to 100 TeV) the barometric coefficient of β = -1,11 %/mb (black line 29) was obtained in [Kozyarivsky, 2004]. In the upper left corner of Fig. 7 a line (nm64) marks the barometric coefficients for the neutron cosmic ray component.

Increase of the absolute values of the barometric coefficients can be explained that single muons are mainly observed up to energies of secondary particles of about 100 GeV. And at higher energies the contribution of muon groups of extensive air showers is more significant. Barometric coefficient for the frequency of extensive air showers depends weakly on the parameters of showers, and it is close to the barometric coefficient for the nucleon component.

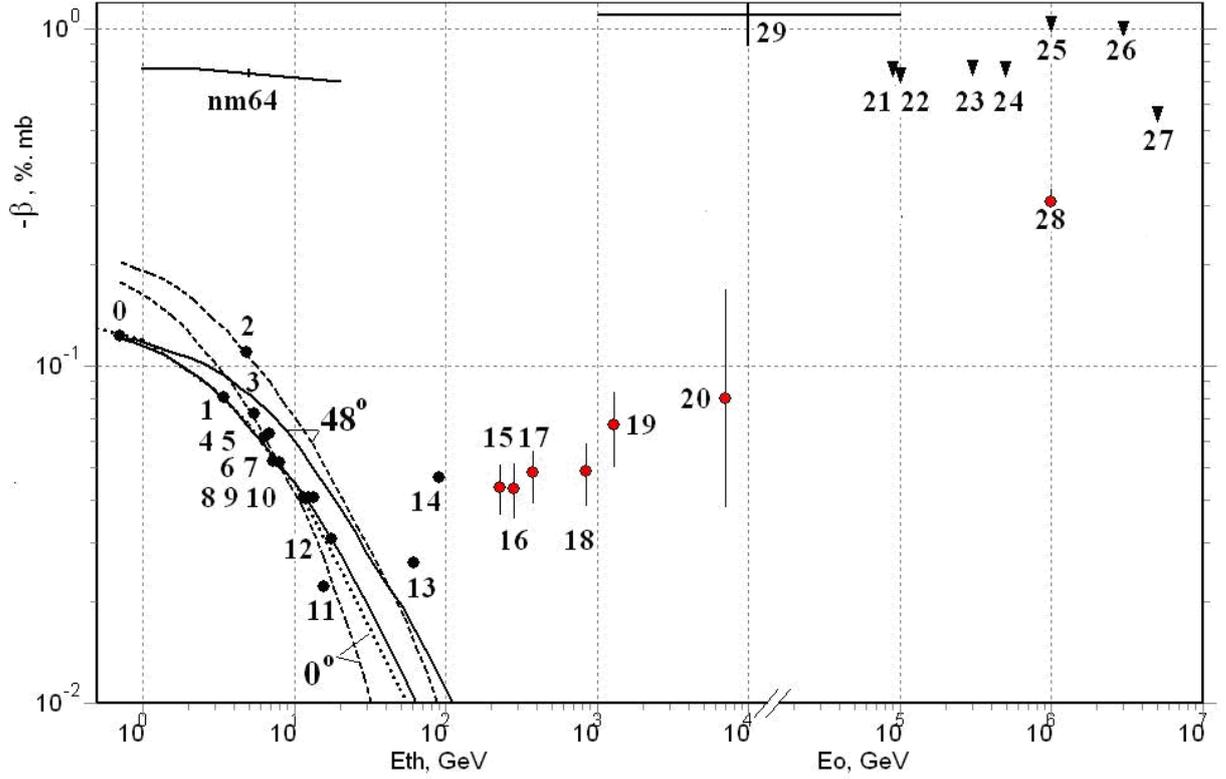

**Fig. 7.** Barometric coefficients for different underground detectors – black points 0-14 [Sagisaka, 1986], and for EAS arrays – points 21-27 [Dorman, 1972]. The solid and dashed curves - calculation for the level of 1013 mb and 600 mb, respectively [Sagisaka, 1986]. The barometric coefficients for the hourly data obtained in this work - red points 15-20. Point 28 – barometric coefficient for group of muons [Tolkacheva, 2011], and point 29 - for the Andyrchi air shower array [Kozyarivsky, 2004]. A line nm64 - the barometric coefficients for the neutron cosmic ray component.

Let us denote intensity of single muons $I_{1\mu}$ and EAS muons $I_{EAS}$, the corresponding barometric coefficients $\beta_{1\mu}$ and $\beta_{EAS}$, the total intensity of single and EAS muons as $I$, and the corresponding barometric coefficient $\beta$. Then we can write down a system of equations in which the first equation determines a sum of relative contributions of the intensities of single and EAS muons, and the second one – a sum and proportion of single and EAS muons in the full barometric coefficient of the detector:

$$\begin{cases} I_{1\mu} + I_{EAS} = I \\ I_{1\mu}/I \cdot \beta_{1\mu} + I_{EAS}/I \cdot \beta_{EAS} = \beta \end{cases} \quad (2)$$

The equation gives a solution for the relative contributions of the two sources of muons:

$$\frac{I_{1\mu}}{I} = \frac{\beta_{EAS} - \beta}{\beta_{EAS} - \beta_{1\mu}} \quad \text{и} \quad \frac{I_{EAS}}{I} = \frac{\beta - \beta_{1\mu}}{\beta_{EAS} - \beta_{1\mu}} \quad (3)$$

The EAS barometric coefficient is $\beta_{EAS}$ = - (0.9 ± 0.1)% / mb. And $\beta_{1\mu}$, $\beta$ and ratio $I_{1\mu}/I$ obtained for Matsushiro, Poatina and the BUST (vertical and 5 intervals of zenith angles) are shown in Table 5.

**Table 5.** The barometric coefficients $\beta_{1\mu}$, $\beta$ and ratio $I_{1\mu}/I$ obtained for Matsushiro, Poatina and the BUST (for the vertical and 5 intervals of zenith angles).

| Detector | Matsu-shiro | Poatina | BUST-v0 $0°<\theta<34°$ | BUST-01 $34°<\theta<48°$ | BUST-02 $48°<\theta<60°$ | BUST-03 $60°<\theta<71°$ | BUST-04 $71°<\theta<80°$ | BUST-05 $80°<\theta<90°$ |
|---|---|---|---|---|---|---|---|---|
| $E_{th}$, GeV | 62 | 91 | 230 | 280 | 380 | 850 | 1300 | >7000 |
| d, m.w.e. | 232 | 365 | 940 | 1170 | 1400 | 1980 | 3260 | 7000 |
| $\beta_{1\mu}$ %/mb | -0.011 | -0.0069 | -0.0025 | -0.0018 | -0.0013 | -0.00065 | -0.00033 | -0.00004 |
| $\beta$ %/mb | -0.0262 ±0.0055 | -0.0469 ±0.0077 | -0.0435 ±0.0075 | -0.0433 ±0.0080 | -0.0485 ±0.0087 | -0.0487 ±0.0102 | -0.0669 ±0.0169 | -0.0798 ±0.0506 |
| $I_{1\mu}/I$, % | 98.3 | 95.5 | 95.4 | 95.4 | 94.7 | 94.6 | 92.6 | 91.1 |

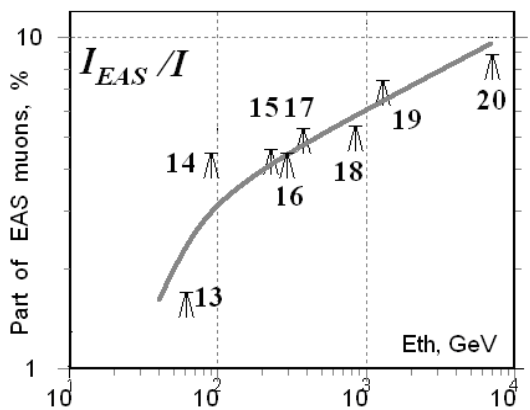

**Fig. 8.** The contribution of EAS muons to the total count rate of the underground detectors.

Using this data the contribution of EAS muons to the total count rate of the underground detectors was depicted in Figure 8.

The results obtained confirm the hypothesis about the contribution of the muon groups to the barometric effect observed according to the BUST. Thus, as expected, the share of muon groups of extensive air showers increases with increasing of the registered particles energy.

After the barometric coefficients were determined, the primary data of the BUST for six intervals of zenith angles have been corrected for the barometric effect according to [Kobelev et al., 2011].

**Comclusions**

The barometric coefficients obtained for the BUST are more than an order of magnitude higher than the theoretically predicted values for these energies.

Barometric effect observed according to the BUST, can be explained by the fact that for the secondary particles with energies > 100 GeV a contribution of muon groups of extensive air showers is becoming more significant. The share of muon groups of extensive air showers increases with the energy of the registered particles. And the share of muon groups of EAS can be estimated by the experimental barometric coefficient for high-energy muons, registered by the underground detectors.

Attempts to explain the large absolute values of the barometric coefficients for underground detectors by the unaccounted temperature effect can be questioned by the fact that the barometric effect is always negative while the temperature effect for high-energy muons is always positive. So, the unaccounted positive temperature effect can hardly affect the increase of the absolute value of the negative barometric effect.

According to the obtained values of the barometric coefficients the primary data of six intervals of zenith angles of the BUST telescope have been corrected for the barometric effect.

**Acknowledgements**
This work is partly supported by Russian FBR grants 11-02-01478 and 12-07-00227, Program № 10 BR of the Presidium RAS "The fundamental properties of matter and Astrophysics.
We thank all the members of the network stations of cosmic rays http://cr0.izmiran.ru/ThankYou.